\documentclass[twocolumn,prl,preprintnumbers,superscriptaddress,landscape,nofootinbib]{revtex4}
\usepackage[utf8]{inputenc}
\usepackage{amsmath}
\usepackage{amsfonts}
\usepackage{amsthm}
\usepackage{float}
\usepackage{braket}
\usepackage{graphicx}
\usepackage{url}
\usepackage{csquotes}
\usepackage[colorlinks=true, pdfstartview=FitV, linkcolor=red, citecolor=blue, urlcolor=blue]{hyperref}
\usepackage{slashed}
\usepackage[normalem]{ulem}

\graphicspath{{./plots/}}

\newcommand{\be}{\begin{equation}}      
\newcommand{\ee}{\end{equation}}      
\newcommand{\bea}{\begin{eqnarray}}      
\newcommand{\eea}{\end{eqnarray}}

\newcommand{\e}{\mathrm{e}}

\newcommand{\Tr}{\mathrm{Tr}}

\begin{document}

\title{Detecting the critical point through entanglement in the Schwinger model
}

\author{Kazuki Ikeda}
\email[]{kazuki.ikeda@stonybrook.edu}
\affiliation{Co-design Center for Quantum Advantage, Stony Brook University, Stony Brook, New York 11794-3800, USA}
\affiliation{Center for Nuclear Theory, Department of Physics and Astronomy, Stony Brook University, Stony Brook, New York 11794-3800, USA}

\author{Dmitri E. Kharzeev}
\email[]{dmitri.kharzeev@stonybrook.edu}
\affiliation{Co-design Center for Quantum Advantage, Stony Brook University, Stony Brook, New York 11794-3800, USA}
\affiliation{Center for Nuclear Theory, Department of Physics and Astronomy, Stony Brook University, Stony Brook, New York 11794-3800, USA}
\affiliation{Department of Physics, Brookhaven National Laboratory, Upton, New York 11973-5000, USA}

\author{Ren\'e Meyer}
\email[]{rene.meyer@uni-wuerzburg.de}
\affiliation{Institute for Theoretical Physics and Astrophysics, Julius--Maximilians--Universität Würzburg, Am Hubland, 97074 Würzburg, Germany}
\affiliation{Würzburg-Dresden Cluster of Excellence ct.qmat}

\author{Shuzhe Shi}
\email[]{shuzhe-shi@tsinghua.edu.cn}
\affiliation{Center for Nuclear Theory, Department of Physics and Astronomy, Stony Brook University, Stony Brook, New York 11794-3800, USA}
\affiliation{Department of Physics, Tsinghua University, Beijing 100084, China}




\bibliographystyle{unsrt}

\begin{abstract}
Using quantum simulations on classical hardware, we study the phase diagram of the massive Schwinger model with a $\theta$-term at finite chemical potential $\mu$. We find that the quantum critical point in the phase diagram of the model can be detected through the entanglement entropy and entanglement spectrum. As a first step, we chart the phase diagram using conventional methods by computing the dependence of the charge and chiral condensates on the fermion mass $m$, coupling constant $g$, and the chemical potential $\mu$. At zero density, the Schwinger model possesses a quantum critical point at $\theta=\pi$ and $m/g \simeq 0.33$. We find that the position of this quantum critical point depends on the chemical potential. Near this quantum critical point, we observe a sharp maximum in the entanglement entropy. Moreover, 
we find that the quantum critical point can be located from the entanglement spectrum by detecting the position of the gap closing point. 
\end{abstract}

\maketitle

\emph{Introduction}. --- 
The Schwinger model \cite{schwinger1962gauge} is quantum electrodynamics in $(1+1)$ space-time dimensions. This model shares many properties with QCD, including confinement (in the massive case), and chiral symmetry breaking. For the case of massless fermions, the Schwinger model is analytically solvable and  equivalent to the theory of a free massive boson field \cite{schwinger1962gauge,brown1963gauge,sommerfield1964definition,zumino1964charge,hagen1967current,lowenstein1971quantum,casher1974vacuum,coleman1975charge,2022arXiv220508860T}. However, the massive Schwinger model cannot be solved analytically, and requires a numerical approach. 

Quantum simulations are potentially advantageous for this purpose, as the Hilbert space grows exponentially with the number of lattice sites. In recent years, there has been a growing interest in quantum algorithms as an efficient, and potentially superior, way to explore the dynamics of quantum field theories, including the Schwinger model~\cite{wallraff2004strong,majer2007coupling,Jordan:2011ne,Jordan:2011ci,Zohar:2012ay,Zohar:2012xf,Banerjee:2012xg,Banerjee:2012pg,Wiese:2013uua,Wiese:2014rla,Jordan:2014tma,Garcia-Alvarez:2014uda,Marcos:2014lda,Bazavov:2015kka,Zohar:2015hwa, Mezzacapo:2015bra, Dalmonte:2016alw, Zohar:2016iic, Martinez:2016yna, Bermudez:2017yrq, gambetta2017building, krinner2018spontaneous, Macridin:2018gdw, Zache:2018jbt, Zhang:2018ufj, Klco:2018kyo, Klco:2018zqz, Lu:2018pjk, Klco:2019xro, Lamm:2018siq, Gustafson:2019mpk, Klco:2019evd, Alexandru:2019ozf, Alexandru:2019nsa, Mueller:2019qqj, Lamm:2019uyc, Magnifico:2019kyj, Chakraborty:2019, Kharzeev:2020kgc, Shaw:2020udc, sahinoslu2020hamiltonian, Paulson:2020zjd, Mathis:2020fuo, Ji:2020kjk, Raychowdhury:2019iki, Davoudi:2020yln, Dasgupta:2020itb, Magnifico:2018wek, PhysRevD.107.L071502}. Another advantage of using quantum algorithms to simulate quantum field theories is the absence of the fermion sign problem at finite chemical potential~\cite{Feynman1982,Preskill:2018fag,Jordan:2012xnu,Preskill:2021apy}. 

In the present work, we use classical hardware to simulate the massive Schwinger model at finite temperature and chemical potential. We analyze the entanglement structure of the ground state, and demonstrate that entanglement can be used as an efficient diagnostic tool for detecting the phase transition, and in particular the presence of the critical point in the phase diagram. 

In particular we find that, at the critical point, there is a level crossing in the entanglement spectrum. Also, near the critical point, the entanglement entropy for a single spatial interval follows predictions from two-dimensional conformal field theory.  It is well known that entanglement is a good probe of topological order~\cite{Jiang2012,PhysRevB.81.064439}. Our results suggests that entanglement can also be used to detect the fluctuating topology as well.

In~\cite{Ikeda:2020agk}, we showed that the presence of the critical point in the phase diagram can be detected through a sharp growth of  topological fluctuations reflected by the real-time topological susceptibility. In this work, we confirm the presence of the critical point at finite and vanishing chemical potential by a calculation of the topological susceptibility, as well as the charge susceptibility.

\if{
The massive Schwinger model undergoes a quantum phase transition at $\theta=\pi$ between the phases with opposite orientations of the electric field. This phase transition is a first-order transition when the mass is sufficiently larger than $g$ $(m\gg g)$. However, Coleman showed \cite{coleman1976more} that the line of the first order phase transition ends at the critical value $m^\ast$, where the phase transition is second order. It is well known that the location of this critical point is at $m^\ast\approx 0.33g$~\cite{schiller1983massive,hamer1982massive,byrnes2002density}. The phase diagram of the theory in the $(m/g, \theta)$ plane is thus reminiscent of the phase diagram of QCD in the $(T,\mu)$ plane of temperature $T$ and baryon chemical potential $\mu$ \cite{rajagopal2001condensed,stephanov2004qcd}. If QCD in 4 dimensions has a critical point at a finite chemical potential, then the lower dimensional model might also exhibit such a critical point. The Schwinger model is a low-dimensional model with properties similar to QCD, which has recently gained popularity in terms of its application to quantum field theory in quantum computers. Therefore, understanding the phase transition of the Schwinger model can be an important step toward realizing the goal of reproducing the phase diagram of QCD using a quantum computer.
}\fi

\vspace{5mm}
\emph{$1+1$-dimensional QED}. --- 
In the presence of finite chemical potential, the action of the massive Schwinger model with $\theta$ term in $(1+1)$-dimensional Minkowski space in temporal gauge $A_0=0$ is
\begin{align} \label{eq:Ham}
H &=\int dz \Big[    
    \frac{E^2}{2}
    -\bar{\psi}(i\gamma^1\partial_1 - g\gamma^1A_1 - m\,\e^{i\gamma_5\theta} 
    + \mu \gamma^0)\psi \Big] \,.
\end{align}
Here, $A_\mu$ is the $U(1)$ gauge potential, $E=\dot{A}_1$ is the corresponding electric field, $\psi$ is a two-component fermion field, $m$ is the fermion mass, $\mu$ is the chemical potential, and $\gamma^\mu$ are two-dimensional $\gamma$-matrices satisfying the Clifford algebra. We use the metric convention $\eta_{\mu\nu}=\mathrm{diag}(1,-1)$.
From Eq.~\eqref{eq:Ham}, we see that the massive theory with a positive mass $m>0$ at $\theta=\pi$ is equivalent to the theory at $\theta=0$ with a negative mass $-m$.

\begin{figure}[!hbtp]
    \centering
    \includegraphics[width=0.95\linewidth]{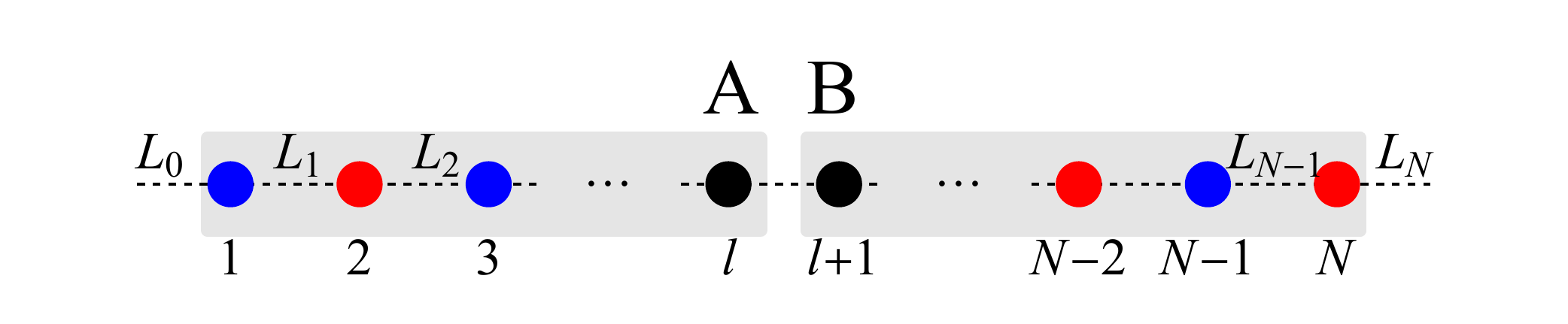}
    \caption{Illustration of the lattice setup. Red dots are fermions and blue dots are anti-fermions. We take the boundary condition as $L_0=-Q/2$ which leads to  $L_{N}=+Q/2$, with $Q$ being the total charge. See text for explanation.
    \label{fig:local_charge}}
\end{figure}

\begin{figure*}[!hbtp]\centering
    \includegraphics[width=\textwidth]{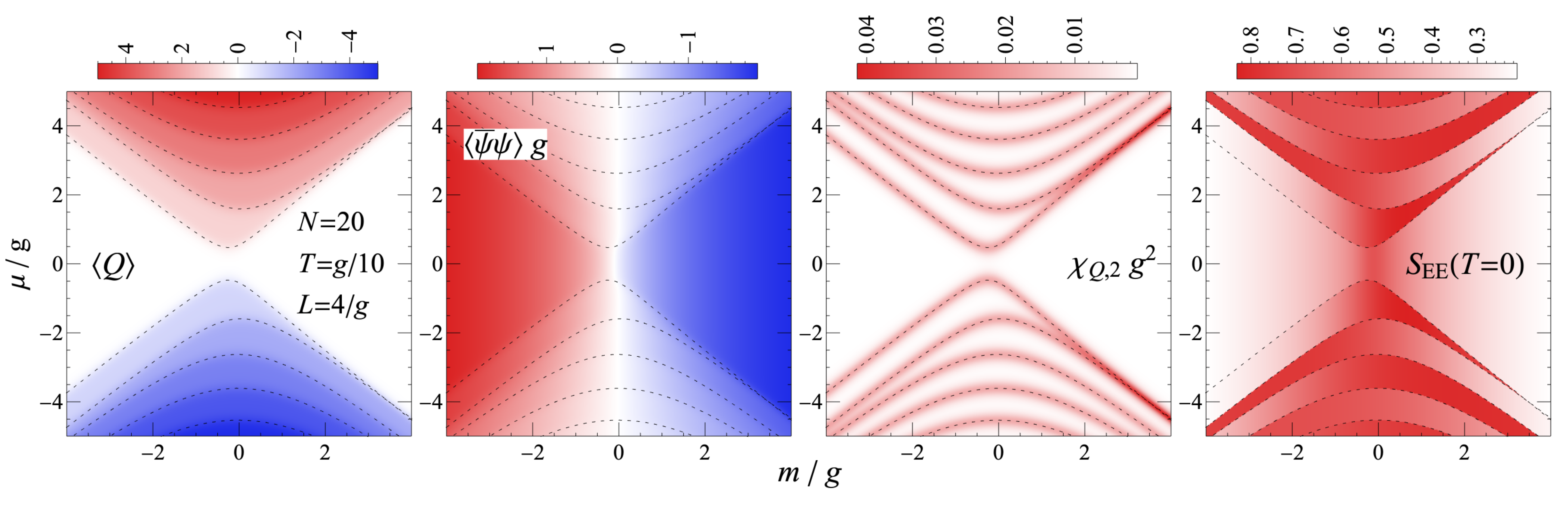}
\caption{\label{fig:diagram}(From left to right) Density plot of total charge, chiral condensate, charge fluctuation, and ground-state left-right entanglement entropy for different values of fermion mass and chemical potential, with $N=20$, $L=4/g$, and $\theta=0$. The entanglement entropy is for the ground state,   corresponding to $T=0$, whereas the other panels are for $T=g/10$. The first-order phase transition line (black dashed contours) are overlaid on top of every panel. See text for more details.}
\end{figure*}

We first discretize the continuum system to obtain the lattice Hamiltonian as follows: We discretize the spatial coordinate as $z=a\,n$, where $a$ is a lattice constant and $n$ is an integer. We choose to work with staggered fermions~\cite{Kogut:1974ag, Susskind:1976jm}, in which a two-component Dirac fermion $\psi=(\psi^1,\psi^2)^T$ is converted to $\psi^1(\psi^2)\to\chi_n/\sqrt{a}$ for odd (even) $n$. We further define $L_n = E(an)/g$ and $U_n = e^{-iagA_1(an)}$, respectively, as the gauge field and gauge link operators connecting the $n^\mathrm{th}$ and $(n+1)^\mathrm{th}$ sites.
We define our Dirac matrices as $\gamma^0 = Z$, $\gamma^1 = i\,Y$, and $\gamma^5 = \gamma^0\gamma^1 =X$, where $X$, $Y$, and $Z$ are the Pauli matrices. The lattice Hamiltonian then reads
\begin{align}
\begin{aligned}
\label{eq:ham2}
H=&\sum_{n=1}^{N-1}
    \big(\frac{im}{2}(-1)^n\sin\theta -\frac{i}{2a} \big)
    \big(U_n\chi^\dagger_{n+1}\chi_{n}-U_n^\dagger\chi^\dagger_{n}\chi_{n+1}\big)\\
&+\sum_{n=1}^{N}\left(m(-1)^n\cos\theta-\mu\right)\chi^\dagger_n\chi_n+\frac{ag^2}{2}\sum_{n=0}^{N}L^2_n.
\end{aligned}
\end{align}

Using open boundary condition ($\chi_0 = \chi_{N+1} = 0$), we can eliminate the link fields $U_n$ by the gauge transformation, $\chi_n\to g_n\chi_n$ and $U_n \to g_{n+1}U_n g_n^\dag$, with $g_1=1$, $g_n = \prod_{m=1}^{n-1}U^\dag_{m}$ (see e.g.~\cite{Ikeda:2020agk, Florio:2023dke}). Gauge invariance of the theory requires that physical states must respect Gauss's law,
\begin{equation} \label{eq:gauss}
    \partial_z E = g\,\bar\psi \gamma^0 \psi. 
\end{equation}
Using~\eqref{eq:gauss}, we get $L_{n}-L_{n-1} = Q_n$, where $Q_n = \chi_n^\dag\chi_n - \frac{1-(-1)^n}{2}$ is the local charge operator. As illustrated in Fig.~\ref{fig:local_charge}, we set the boundary electric field as $L_0=-Q/2$, with $Q\equiv \sum_{n=1}^{N} Q_n$ being the total charge. This leads to the solution 
\begin{align}
    L_n=-\frac{Q}{2}+\sum_{m=1}^n Q_m\,,
    \label{eq:gauss_solution}
\end{align}
In this setup, the charge conjugation (C) is equivalent to the parity reflection (P). The P parity is explicitly preserved at $\theta =0$, and so is C conjugation.

Finally, we use the Jordan--Wigner transformation~\cite{Jordan:1928wi} $\chi_n=\frac{X_n-iY_n}{2}\prod_{m=1}^{n-1}(-i Z_m)$, after which the Hamiltonian becomes
\begin{align}
\begin{aligned}
\label{eq:Hamiltonian}
H=&\sum_{n=1}^{N-1}
    \left(\frac{1}{4a}-\frac{m}{4}(-1)^n\sin\theta\right)
    \left(X_n X_{n+1}+Y_n Y_{n+1}\right)\\
&+\sum_{n=1}^{N}\frac{m(-1)^n\cos\theta-\mu}{2} Z_n+\frac{ag^2}{2}\sum_{n=0}^{N}L^2_n.
\end{aligned}
\end{align}
In our study, we diagonalize this Hamiltonian to obtain eigenstates and compute expectation values. 

\vspace{5mm}
\emph{Phase Diagram}. --- 
The expectation value of an observable $O$ at finite temperature is defined by  $\langle O \rangle \equiv \Tr( \rho_\mathrm{th}\, O )$, where the thermal density matrix
\begin{align}
    \rho_\mathrm{th} \equiv \frac{e^{- H/T}}{\Tr (e^{- H/T})}
\end{align}
is a function of temperature ($T$), chemical potential, and mass. Observables of interest to us include the total charge $(\langle Q \rangle\equiv\int\langle\psi^\dagger\psi\rangle\mathrm{d}z)$, chiral condensate ($\langle\bar{\psi}\psi\rangle \equiv {\int \langle \bar{\psi}\psi\rangle dz}/L$), and charge susceptibility ($\chi_{Q,2} \equiv (\langle Q \rangle^2 - \langle Q^2 \rangle)/L^2$), with $L = N\,a$ being the lattice size.

In particular, we study the entanglement entropy between two subsystems $A = \{1,2,\cdots,\ell\}$ and $B=\{\ell+1, \cdots, N\}$, see Fig.~\ref{fig:local_charge}, with $A\cap B=\emptyset, A\cup B=\{1,2,\cdots,N\}$. 
Let $\rho$ be a density operator of the entire system $A\cup B$. The entanglement entropy between $A$ and $B$ is defined by
$S_{EE,\ell} \equiv -\Tr_{A}(\rho_A\log\rho_A)$, where $\rho_A$ is defined by tracing out the Hilbert space of $B$: $\rho_A=\Tr_{B}\rho$. The entanglement spectrum, $\{\lambda_i\}$, is spanned by the eigenvalues of $\rho_A$. In this study we choose $\rho$ as the ground state of the Hamiltonian, $\rho=\ket{\psi_{gs}}\bra{\psi_{gs}}$.

In Fig.~\ref{fig:diagram}, we show our results for the thermal expectation values of the charge, chiral condensate, charge fluctuation, and ground-state  entanglement entropy ($T=0$ and $\ell = N/2$) for different values of fermion mass and chemical potential at $T=g/10$ and $\theta=0$, with lattice parameters $N=10$, $a=0.5/g$,  
Under the charge conjugation $\mu \to -\mu$, we confirm that 
the total charge is C-odd, $\langle Q\rangle \to -\langle Q\rangle$, whereas the chiral condensate, second-order charge susceptibility, and entanglement entropy are C-even.

We note that at finite chemical potential, the Hamiltonian~\eqref{eq:Hamiltonian} can be written as $H(\mu) = H_0 - \mu\,Q$, where the total charge $Q$ commutes with $H_0 \equiv H(\mu=0)$ and, therefore, is a good quantum number. Thus, we may label a quantum state by its energy level and electric charge, $\ket{q,n}$, representing the $n^\mathrm{th}$ lowest energy eigenstate that carries $q$ units of charge.
We also denote that $ H_0 \ket{q,n} = E_{n}^{(q)}\ket{q,n}$ and $ Q\ket{q,n} = q\ket{q,n}$. Charge conjugation symmetry requires that $E_{n}^{(-q)} = E_{n}^{(q)}$. At vanishing chemical potential, we find that the $q$-dependent lowest energies follow a hierarchy of charge $E_{0}^{(q=0)} < E_{0}^{(|q|=1)} < E_{0}^{(|q|=2)} < \cdots $, and the gaps between the neighbouring energy levels, $E_{0}^{(|q|+1)} - E_{0}^{(|q|)}$, increase with $|q|$. This is caused by the Coulomb repulsive interaction between charged fermions. At finite $\mu$, the energy would be shifted according to the charge, $H(\mu)\ket{q,n} = (E_{n}^{(q)} - \mu\, q)\ket{q,n}$. When $\mu$ crosses $E_{n}^{(q=1)}-E_{n}^{(q=0)}$ from below, the ground state of the system switches from $q=0$ to $q=1$. By a similar argument, we find the charge of the ground state to be $q$ in the interval $(E_{n}^{(q)}-E_{n}^{(q-1)}) < \mu < (E_{n}^{(q+1)}-E_{n}^{(q)})$.

\begin{figure}[!hbtp]
\centering
    \includegraphics[width=0.4\textwidth]{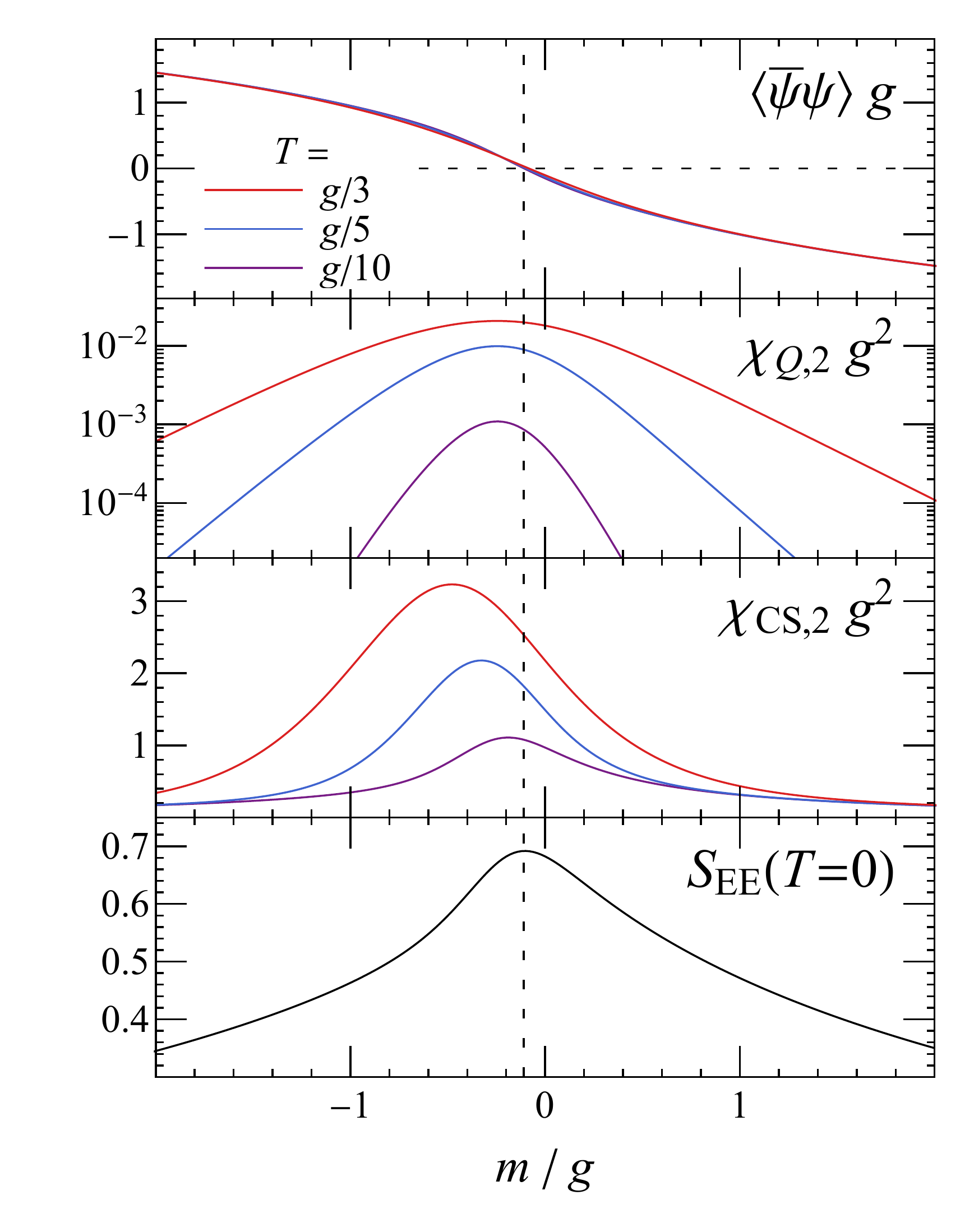}
    \caption{(From top to bottom) Fermion mass dependence of chiral condensate, vector charge susceptibility, chiral susceptibility, and vacuum entanglement entropy, with $\theta=0$ and $\mu=0$. In the top three panels, red, blue, and purple curves represent $T=g/3$, $g/5$, and $g/10$, respectively.}
    \label{fig:susceptibility}
\end{figure}

These discrete energy levels depend on the fermion mass $m$, and define the lines of the first-order phase transitions in the phase diagram.  These boundaries are shown by black dashed curves in all panels of Fig.~\ref{fig:diagram}. These boundaries separate different ground states, and we observe different values of entanglement entropy in each state. At low temperature, $T=g/10$, we observe a clear jump of the total charge across the boundaries, where the charge susceptibility also peaks.

Motivated by the peak in the entanglement entropy in the branch around $\mu=0$, we investigate the phase with vanishing chemical potential in more detail. We show the entanglement entropy, the scalar condensate, and the charge and topological susceptibilities at $\mu=0$ in Fig.~\ref{fig:susceptibility}. 
The topological susceptibility is defined as $\chi_{\mathrm{CS},2} = (\langle q_\mathrm{CS}^2\rangle - \langle q_\mathrm{CS}\rangle^2)$, where
$q_\mathrm{CS}=\partial_\mu K^\mu$ is the Chern--Pontryagin number density associated with the divergence of the Chern--Simons current ($K^\mu$). In $(1+1)$ dimensions, $K^\mu = (g/2\pi)\varepsilon^{\mu\nu} A_\nu$ and $q_\mathrm{CS}=(g/4\pi)\varepsilon^{\mu\nu}F_{\mu\nu} = (g/2\pi)E  = (g^2/2\pi) \sum_{n=1}^{N-1} L_n/(N-1)$. 

We observe that entanglement entropy peaks at $m^*\approx-0.11\,g$, where the chiral condensate also vanishes. Meanwhile, fluctuations of both vector charge and Chern--Simons numbers peak around the same mass $m^*$ at low temperatures. 
The value of $m^*$ is consistent with previous calculation~\cite{Ikeda:2020agk}, which takes the real-time topological susceptibility to detect the critical point.

The coincidence of the critical point of the phase diagram with the peak of the entanglement entropy is known for some models. For example, it is suggested that entanglement measures are suitable tools for analyzing phase transitions in spin systems~\cite{Filippone:2011aa, PhysRevA.95.042321}. However, as stressed in~\cite{Filippone:2011aa}, the discontinuity in entanglement entropy is not always related to a critical point in a phase diagram. 

\begin{figure*}
    \centering
    \includegraphics[width=0.8\textwidth]
    {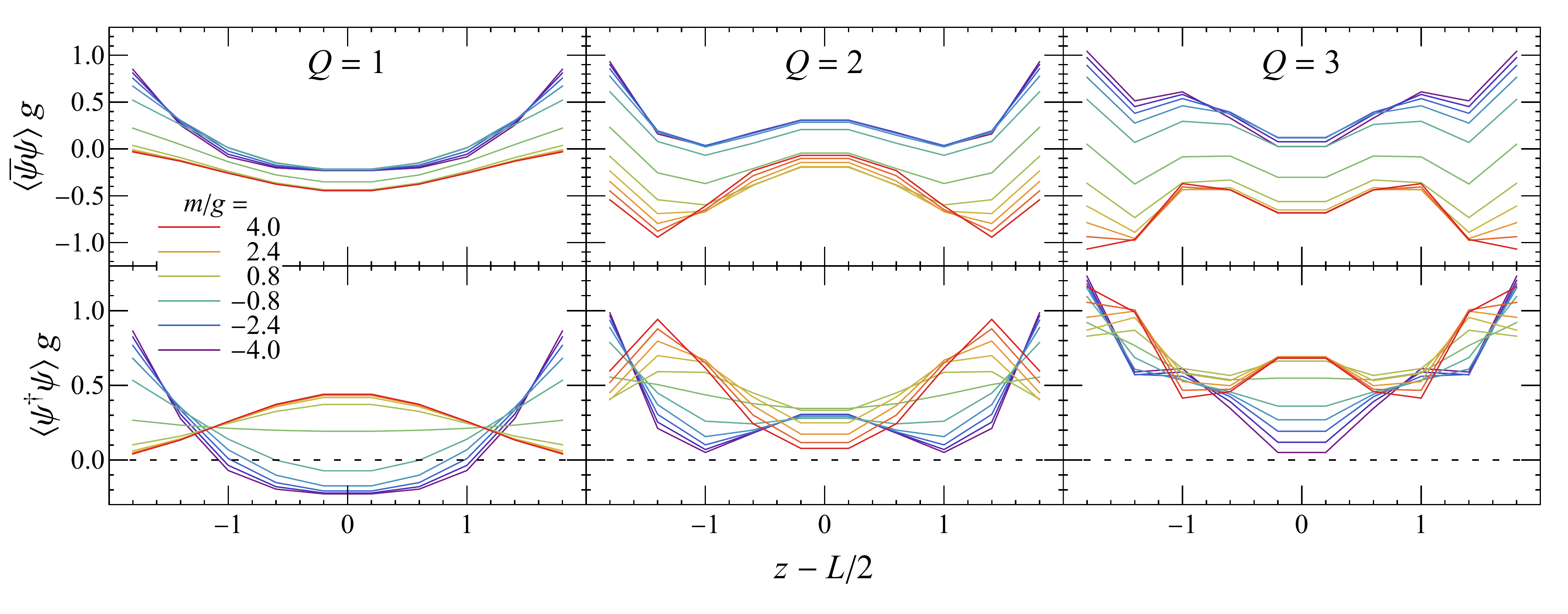}
    \caption{(From left to right) Spatial distribution of chiral condensate (upper panel) and charge density (lower panel) for $Q=1$, $2$, and $3$ states, with shown values of fermion mass.}
    \label{fig:crystal}
\end{figure*}
In the $\mu\neq 0$ branches of the phase diagram in Fig.~\ref{fig:diagram}, we observe that entanglement entropy monotonically increases (decreases) with mass for odd (even) charge. This is due to the mass dependence of the charge density in the middle of system, at the boundary of subsystems $A$ and $B$, see lower panels in Fig.~\ref{fig:crystal}. For finite chemical potential, we see a hint of the crystalline structure discussed in~\cite{Fischler:1978ms, Kao:1994wv, Nagy:2004ey} in Fig.~\ref{fig:crystal}.  

\begin{figure}[!htbp]
    \centering
    \includegraphics[width=0.4\textwidth]{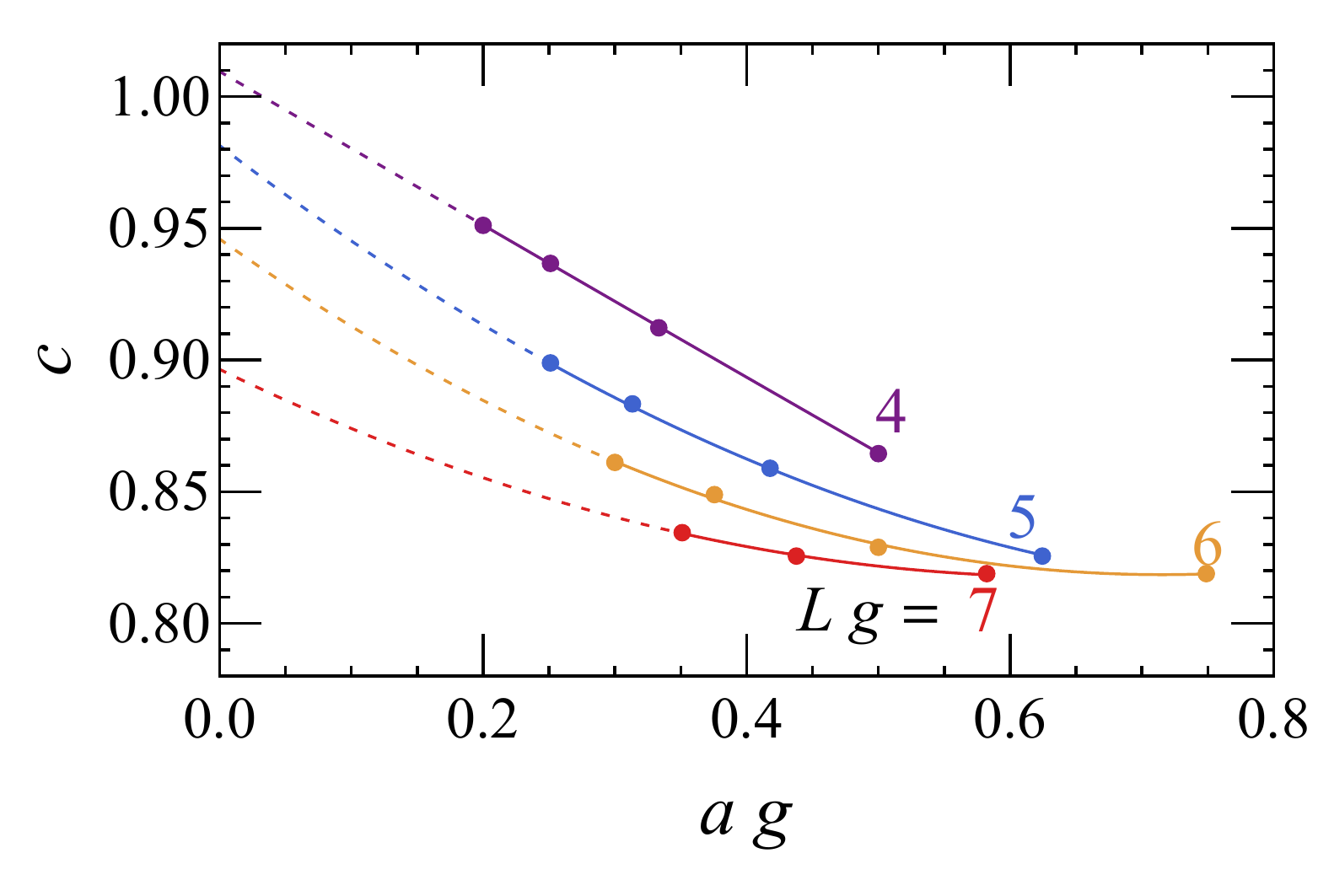}
    \caption{Central charge versus lattice spacing. The dashed curves are second order polynomial fits to guide the eye. They come close to the expected central charge $c=1$ in the continuum limit $a\rightarrow 0$ within 15\% accuracy.
    Results are evaluated at vanishing chemical potential.}
    \label{fig:cc}
\end{figure}
In a conformal field theory on a finite interval of length $L=a N$, the entanglement entropy of a subsystem with length $\ell = a l$ and its complement obeys~\cite{Calabrese:2009qy} 
\begin{align}
    S_{EE,l} = \frac{c}{6}\ln\Big(\frac{2N}{\pi}\sin\frac{\pi\,l}{N}\Big) - c'\,,
    \label{eq:SEE_l}
\end{align}
where $c$ is the central charge. It is known that the massive Schwinger model at critical mass $m^*$ is a conformal fixed point and belongs to the universality class of the classical 2d Ising model, in which the central charge is expected to approach $c=1$ in the continuum limit. In what follows we study the entanglement entropy between the subsystems $A$ and $B$ are defined as in Fig.~\ref{fig:local_charge}, where the length of $A$ is taken to be $\ell$. 
We study the finite size effects in the entanglement entropy by varying the lattice spacing ($a$) and number of lattice sites ($N$). For each choice of $a$ and $N$, we first determine the critical mass by maximizing $S_{EE}$ when $l=N/2$ and then compute $S_{EE,l}$ at different values of $l$. We refer to Fig.~\ref{fig:EE} of the Supplemental Material for more details. By fitting the results using Eq.~\eqref{eq:SEE_l}, we obtain the central charge (see Fig.~\ref{fig:cc}).
It is clear that the $l$ dependence of $S_{EE,l}$ is well-described by~\eqref{eq:SEE_l}, and the corresponding central charge is within $\sim 15\%$ from $c=1$. We suspect that the deviation from $c=1$ is caused by finite size effects, as the continuum limit $ag\rightarrow0$ in Fig.~\ref{fig:cc} is beyond the reach of our numerical capacity. This is supported by the fact that $c$ seems to approach unity when the lattice spacing decreases and the number of sites increases.

\begin{figure}
    \centering
    \includegraphics[width=0.4\textwidth]
    {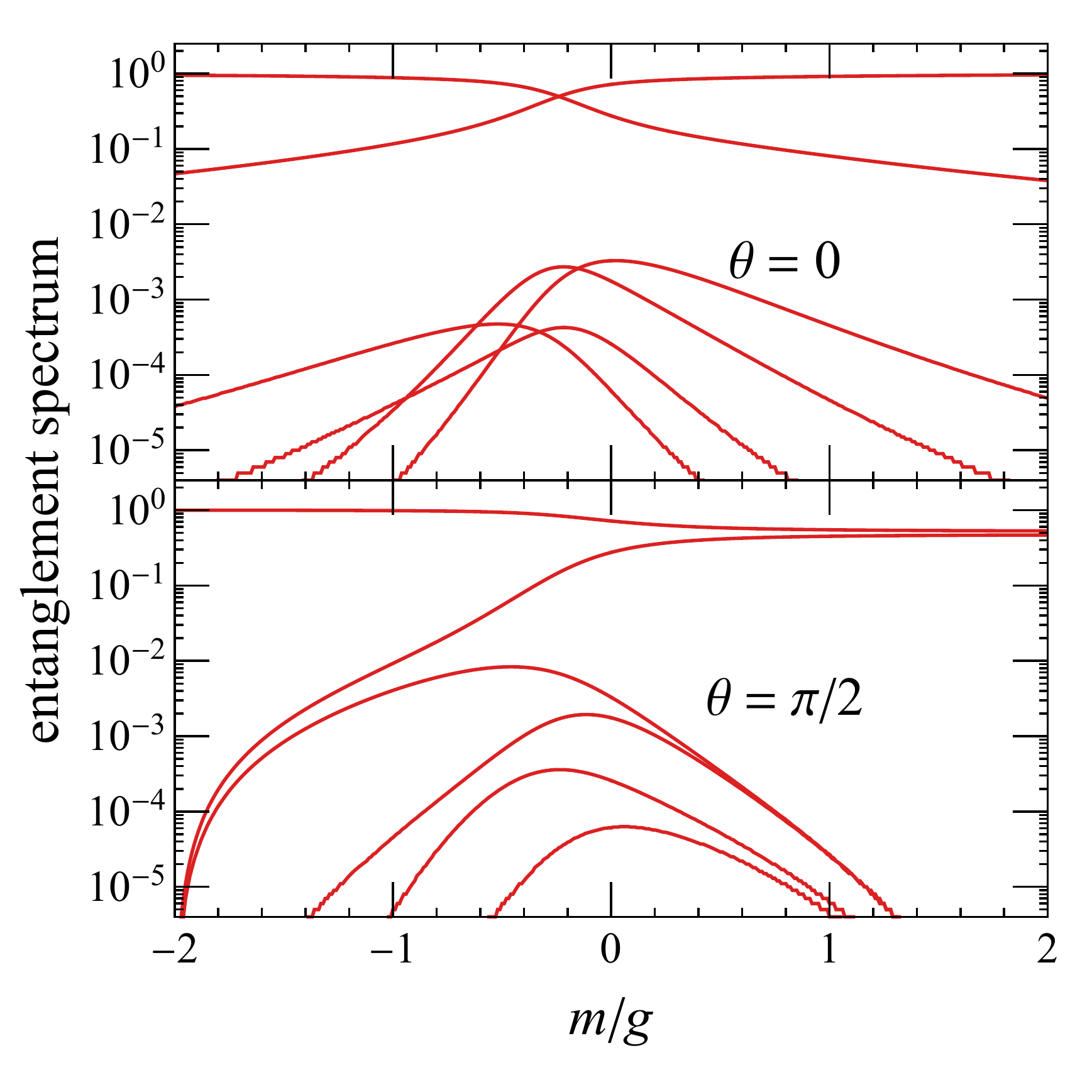}
    \caption{Entanglement spectrum at $\theta=0$ (upper panel) and $\theta=\pi/2$ (lower panel). Other parameters are set to $N=10$, $a=0.5/g$ and $\mu=0$. }
    \label{fig:ES_gap}
\end{figure}
Finally, we study the entanglement spectrum. Suppose that the dimension of the Hilbert space of subsystem $A$ is not larger than that of subsystem $B$: $\text{dim}\mathcal{H}_A\le\text{dim}\mathcal{H}_B$. The Gram--Schmidt orthonormalization of a quantum state is
\begin{equation}
    \ket{\psi} = \sum_{\alpha=1}^{\text{dim}\mathcal{H}_A}\xi_\alpha\ket{\psi^A_\alpha}_A\ket{\psi^B_\alpha}_B ,
\end{equation}
and the coefficients $\xi_\alpha~(\alpha=1,\cdots,\text{dim} \mathcal{H}_A)$ constitute the entanglement spectrum~\cite{PhysRevB.81.064439}. 

Our simulations show that the gap in the entanglement spectrum closes at the critical point, i.e. at $m^* \simeq -0.11 g$, see the upper panel of Fig.~\ref{fig:ES_gap}. We present the results at $\mu=0$, but we find similar band crossing points in the entanglement spectrum at finite chemical potential as well. The parity symmetry in the Schwinger model is broken if $\theta\neq0,\pi$ due to the background electric field, which results in a non-zero gap in the entanglement spectrum.  We confirm this by showing in the lower panel of Fig.~\ref{fig:ES_gap} the entanglement spectrum at $\theta = \pi/2$.

To summarize, we have performed a quantum simulation of the massive Schwinger model at finite chemical potential and temperature, and have found that the location of the critical point in the phase diagram can be determined from the entanglement entropy and the entanglement spectrum. The entanglement entropy peaks in the vicinity of the critical point, showing conformal behavior.  The entanglement spectrum near the critical point exhibits the level crossing phenomenon. We speculate that the entanglement entropy and entanglement spectrum show a similar behavior near the critical points in other, more complicated field theories, including QCD. It would be interesting to confirm these findings in other theories such as holographic QCD \cite{Sakai:2004cn,yee2011holographic}, and to investigate whether the symmetry-resolved entanglement \cite{goldstein2018symmetry} in the Schwinger model can indicate the presence of the critical point as well. 

\section*{Acknowledgement}
We thank Giuseppe Di Giulio and Vladimir Korepin for useful discussions. This work was supported by the U.S. Department of Energy, Office of Science, National Quantum Information Science Research Centers, Co-design Center for Quantum Advantage (C2QA) under Contract No.DE-SC0012704 (KI), and the U.S. Department of Energy, Office of Science, Office of Nuclear Physics, Grants Nos. DE-FG88ER41450 (DK, SS) and DE-SC0012704 (DK). The work of R.M. was 
supported by the Deutsche Forschungsgemeinschaft (DFG, German Research Foundation) through Project-ID 258499086-SFB 1170 \enquote{ToCoTronics} and through the W\"{u}rzburg-Dresden Cluster of Excellence on Complexity and Topology in Quantum Matter – ct.qmat Project-ID 
390858490—EXC 2147. 

\bibliographystyle{utphys}
\bibliography{ref}

\providecommand{\href}[2]{#2}\begingroup\raggedright\begin{thebibliography}{10}

\bibitem{schwinger1962gauge}
J.~S. Schwinger, ``{Gauge Invariance and Mass},''
  \href{http://dx.doi.org/10.1103/PhysRev.125.397}{{\em Phys. Rev.} {\bfseries
  125} (1962) 397--398}.
[,151(1962)].

\bibitem{brown1963gauge}
L.~Brown, ``Gauge invariance and mass in a two-dimensional model,'' {\em Il
  Nuovo Cimento (1955-1965)} {\bfseries 29} no.~3, (1963) 617--643.

\bibitem{sommerfield1964definition}
C.~M. Sommerfield, ``On the definition of currents and the action principle in
  field theories of one spatial dimension,'' {\em Annals of Physics} {\bfseries
  26} no.~1, (1964) 1--43.

\bibitem{zumino1964charge}
B.~Zumino, ``Charge conservation and the mass of the photon,'' {\em Phys.
  Lett.} {\bfseries 10} no.~CERN-TH-425, (1964) 224--226.

\bibitem{hagen1967current}
C.~Hagen, ``Current definition and mass renormalization in a model field
  theory,'' {\em Il Nuovo Cimento A (1965-1970)} {\bfseries 51} no.~4, (1967)
  1033--1052.

\bibitem{lowenstein1971quantum}
J.~Lowenstein and J.~Swieca, ``Quantum electrodynamics in two dimensions,''
  {\em Annals of Physics} {\bfseries 68} no.~1, (1971) 172--195.

\bibitem{casher1974vacuum}
A.~Casher, J.~Kogut, and L.~Susskind, ``Vacuum polarization and the absence of
  free quarks,'' {\em Physical Review D} {\bfseries 10} no.~2, (1974) 732.

\bibitem{coleman1975charge}
S.~Coleman, R.~Jackiw, and L.~Susskind, ``Charge shielding and quark
  confinement in the massive schwinger model,'' {\em Annals of Physics}
  {\bfseries 93} no.~1-2, (1975) 267--275.

\bibitem{2022arXiv220508860T}
A.~{Tomiya}, ``{Schwinger model at finite temperature and density with beta
  VQE},'' {\em arXiv e-prints} (May, 2022) arXiv:2205.08860,
  \href{http://arxiv.org/abs/2205.08860}{{\ttfamily arXiv:2205.08860
  [hep-lat]}}.

\bibitem{wallraff2004strong}
A.~Wallraff, D.~I. Schuster, A.~Blais, L.~Frunzio, R.~S. Huang, J.~Majer,
  S.~Kumar, S.~M. Girvin, and R.~J. Schoelkopf, ``Strong coupling of a single
  photon to a superconducting qubit using circuit quantum electrodynamics,''
  \href{http://dx.doi.org/10.1038/nature02851}{{\em Nature} {\bfseries 431}
  no.~7005, (2004) 162--167}.

\bibitem{majer2007coupling}
J.~Majer, J.~M. Chow, J.~M. Gambetta, J.~Koch, B.~R. Johnson, J.~A. Schreier,
  L.~Frunzio, D.~I. Schuster, A.~A. Houck, A.~Wallraff, A.~Blais, M.~H.
  Devoret, S.~M. Girvin, and R.~J. Schoelkopf, ``Coupling superconducting
  qubits via a cavity bus,'' \href{http://dx.doi.org/10.1038/nature06184}{{\em
  Nature} {\bfseries 449} no.~7161, (2007) 443--447}.

\bibitem{Jordan:2011ne}
S.~P. Jordan, K.~S.~M. Lee, and J.~Preskill, ``{Quantum Algorithms for Quantum
  Field Theories},'' \href{http://dx.doi.org/10.1126/science.1217069}{{\em
  Science} {\bfseries 336} (2012) 1130--1133},
\href{http://arxiv.org/abs/1111.3633}{{\ttfamily arXiv:1111.3633 [quant-ph]}}.

\bibitem{Jordan:2011ci}
S.~P. Jordan, K.~S.~M. Lee, and J.~Preskill, ``{Quantum Computation of
  Scattering in Scalar Quantum Field Theories},''
  \href{http://arxiv.org/abs/1112.4833}{{\ttfamily arXiv:1112.4833 [hep-th]}}.
[Quant. Inf. Comput.14,1014(2014)].

\bibitem{Zohar:2012ay}
E.~Zohar, J.~I. Cirac, and B.~Reznik, ``{Simulating Compact Quantum
  Electrodynamics with ultracold atoms: Probing confinement and nonperturbative
  effects},'' \href{http://dx.doi.org/10.1103/PhysRevLett.109.125302}{{\em
  Phys. Rev. Lett.} {\bfseries 109} (2012) 125302},
\href{http://arxiv.org/abs/1204.6574}{{\ttfamily arXiv:1204.6574 [quant-ph]}}.

\bibitem{Zohar:2012xf}
E.~Zohar, J.~I. Cirac, and B.~Reznik, ``{Cold-Atom Quantum Simulator for SU(2)
  Yang-Mills Lattice Gauge Theory},''
  \href{http://dx.doi.org/10.1103/PhysRevLett.110.125304}{{\em Phys. Rev.
  Lett.} {\bfseries 110} no.~12, (2013) 125304},
\href{http://arxiv.org/abs/1211.2241}{{\ttfamily arXiv:1211.2241 [quant-ph]}}.

\bibitem{Banerjee:2012xg}
D.~Banerjee, M.~Bögli, M.~Dalmonte, E.~Rico, P.~Stebler, U.~J. Wiese, and
  P.~Zoller, ``{Atomic Quantum Simulation of U(N) and SU(N) Non-Abelian Lattice
  Gauge Theories},''
  \href{http://dx.doi.org/10.1103/PhysRevLett.110.125303}{{\em Phys. Rev.
  Lett.} {\bfseries 110} no.~12, (2013) 125303},
\href{http://arxiv.org/abs/1211.2242}{{\ttfamily arXiv:1211.2242
  [cond-mat.quant-gas]}}.

\bibitem{Banerjee:2012pg}
D.~Banerjee, M.~Dalmonte, M.~Muller, E.~Rico, P.~Stebler, U.~J. Wiese, and
  P.~Zoller, ``{Atomic Quantum Simulation of Dynamical Gauge Fields coupled to
  Fermionic Matter: From String Breaking to Evolution after a Quench},''
  \href{http://dx.doi.org/10.1103/PhysRevLett.109.175302}{{\em Phys. Rev.
  Lett.} {\bfseries 109} (2012) 175302},
\href{http://arxiv.org/abs/1205.6366}{{\ttfamily arXiv:1205.6366
  [cond-mat.quant-gas]}}.

\bibitem{Wiese:2013uua}
U.-J. Wiese, ``{Ultracold Quantum Gases and Lattice Systems: Quantum Simulation
  of Lattice Gauge Theories},''
  \href{http://dx.doi.org/10.1002/andp.201300104}{{\em Annalen Phys.}
  {\bfseries 525} (2013) 777--796},
\href{http://arxiv.org/abs/1305.1602}{{\ttfamily arXiv:1305.1602 [quant-ph]}}.

\bibitem{Wiese:2014rla}
U.-J. Wiese, ``{Towards Quantum Simulating QCD},''
  \href{http://dx.doi.org/10.1016/j.nuclphysa.2014.09.102}{{\em Nucl. Phys.}
  {\bfseries A931} (2014) 246--256},
\href{http://arxiv.org/abs/1409.7414}{{\ttfamily arXiv:1409.7414 [hep-th]}}.

\bibitem{Jordan:2014tma}
S.~P. Jordan, K.~S.~M. Lee, and J.~Preskill, ``{Quantum Algorithms for
  Fermionic Quantum Field Theories},''
\href{http://arxiv.org/abs/1404.7115}{{\ttfamily arXiv:1404.7115 [hep-th]}}.

\bibitem{Garcia-Alvarez:2014uda}
L.~García-Álvarez, J.~Casanova, A.~Mezzacapo, I.~L. Egusquiza, L.~Lamata,
  G.~Romero, and E.~Solano, ``{Fermion-Fermion Scattering in Quantum Field
  Theory with Superconducting Circuits},''
  \href{http://dx.doi.org/10.1103/PhysRevLett.114.070502}{{\em Phys. Rev.
  Lett.} {\bfseries 114} no.~7, (2015) 070502},
\href{http://arxiv.org/abs/1404.2868}{{\ttfamily arXiv:1404.2868 [quant-ph]}}.

\bibitem{Marcos:2014lda}
D.~Marcos, P.~Widmer, E.~Rico, M.~Hafezi, P.~Rabl, U.~J. Wiese, and P.~Zoller,
  ``{Two-dimensional Lattice Gauge Theories with Superconducting Quantum
  Circuits},'' \href{http://dx.doi.org/10.1016/j.aop.2014.09.011}{{\em Annals
  Phys.} {\bfseries 351} (2014) 634--654},
\href{http://arxiv.org/abs/1407.6066}{{\ttfamily arXiv:1407.6066 [quant-ph]}}.

\bibitem{Bazavov:2015kka}
A.~Bazavov, Y.~Meurice, S.-W. Tsai, J.~Unmuth-Yockey, and J.~Zhang,
  ``{Gauge-invariant implementation of the Abelian Higgs model on optical
  lattices},'' \href{http://dx.doi.org/10.1103/PhysRevD.92.076003}{{\em Phys.
  Rev.} {\bfseries D92} no.~7, (2015) 076003},
\href{http://arxiv.org/abs/1503.08354}{{\ttfamily arXiv:1503.08354 [hep-lat]}}.

\bibitem{Zohar:2015hwa}
E.~Zohar, J.~I. Cirac, and B.~Reznik, ``{Quantum Simulations of Lattice Gauge
  Theories using Ultracold Atoms in Optical Lattices},''
  \href{http://dx.doi.org/10.1088/0034-4885/79/1/014401}{{\em Rept. Prog.
  Phys.} {\bfseries 79} no.~1, (2016) 014401},
\href{http://arxiv.org/abs/1503.02312}{{\ttfamily arXiv:1503.02312
  [quant-ph]}}.

\bibitem{Mezzacapo:2015bra}
A.~Mezzacapo, E.~Rico, C.~Sabín, I.~L. Egusquiza, L.~Lamata, and E.~Solano,
  ``{Non-Abelian $SU(2)$ Lattice Gauge Theories in Superconducting Circuits},''
  \href{http://dx.doi.org/10.1103/PhysRevLett.115.240502}{{\em Phys. Rev.
  Lett.} {\bfseries 115} no.~24, (2015) 240502},
\href{http://arxiv.org/abs/1505.04720}{{\ttfamily arXiv:1505.04720
  [quant-ph]}}.

\bibitem{Dalmonte:2016alw}
M.~Dalmonte and S.~Montangero, ``{Lattice gauge theory simulations in the
  quantum information era},''
  \href{http://dx.doi.org/10.1080/00107514.2016.1151199}{{\em Contemp. Phys.}
  {\bfseries 57} no.~3, (2016) 388--412},
\href{http://arxiv.org/abs/1602.03776}{{\ttfamily arXiv:1602.03776
  [cond-mat.quant-gas]}}.

\bibitem{Zohar:2016iic}
E.~Zohar, A.~Farace, B.~Reznik, and J.~I. Cirac, ``{Digital lattice gauge
  theories},'' \href{http://dx.doi.org/10.1103/PhysRevA.95.023604}{{\em Phys.
  Rev.} {\bfseries A95} no.~2, (2017) 023604},
\href{http://arxiv.org/abs/1607.08121}{{\ttfamily arXiv:1607.08121
  [quant-ph]}}.

\bibitem{Martinez:2016yna}
E.~A. Martinez {\em et~al.}, ``{Real-time dynamics of lattice gauge theories
  with a few-qubit quantum computer},''
  \href{http://dx.doi.org/10.1038/nature18318}{{\em Nature} {\bfseries 534}
  (2016) 516--519},
\href{http://arxiv.org/abs/1605.04570}{{\ttfamily arXiv:1605.04570
  [quant-ph]}}.

\bibitem{Bermudez:2017yrq}
A.~Bermudez, G.~Aarts, and M.~Müller, ``{Quantum sensors for the generating
  functional of interacting quantum field theories},''
  \href{http://dx.doi.org/10.1103/PhysRevX.7.041012}{{\em Phys. Rev.}
  {\bfseries X7} no.~4, (2017) 041012},
\href{http://arxiv.org/abs/1704.02877}{{\ttfamily arXiv:1704.02877
  [quant-ph]}}.

\bibitem{gambetta2017building}
J.~M. Gambetta, J.~M. Chow, and M.~Steffen, ``Building logical qubits in a
  superconducting quantum computing system,''
  \href{http://dx.doi.org/10.1038/s41534-016-0004-0}{{\em npj Quantum
  Information} {\bfseries 3} no.~1, (2017) 2}.

\bibitem{krinner2018spontaneous}
L.~Krinner, M.~Stewart, A.~Pazmi{\~n}o, J.~Kwon, and D.~Schneble, ``Spontaneous
  emission of matter waves from a tunable open quantum system,''
  \href{http://dx.doi.org/10.1038/s41586-018-0348-z}{{\em Nature} {\bfseries
  559} no.~7715, (2018) 589--592}.

\bibitem{Macridin:2018gdw}
A.~Macridin, P.~Spentzouris, J.~Amundson, and R.~Harnik, ``{Electron-Phonon
  Systems on a Universal Quantum Computer},''
  \href{http://dx.doi.org/10.1103/PhysRevLett.121.110504}{{\em Phys. Rev.
  Lett.} {\bfseries 121} no.~11, (2018) 110504},
\href{http://arxiv.org/abs/1802.07347}{{\ttfamily arXiv:1802.07347
  [quant-ph]}}.

\bibitem{Zache:2018jbt}
T.~V. Zache, F.~Hebenstreit, F.~Jendrzejewski, M.~K. Oberthaler, J.~Berges, and
  P.~Hauke, ``{Quantum simulation of lattice gauge theories using Wilson
  fermions},'' \href{http://dx.doi.org/10.1088/2058-9565/aac33b}{{\em Sci.
  Technol.} {\bfseries 3} (2018) 034010},
\href{http://arxiv.org/abs/1802.06704}{{\ttfamily arXiv:1802.06704
  [cond-mat.quant-gas]}}.

\bibitem{Zhang:2018ufj}
J.~Zhang, J.~Unmuth-Yockey, J.~Zeiher, A.~Bazavov, S.~W. Tsai, and Y.~Meurice,
  ``{Quantum simulation of the universal features of the Polyakov loop},''
  \href{http://dx.doi.org/10.1103/PhysRevLett.121.223201}{{\em Phys. Rev.
  Lett.} {\bfseries 121} no.~22, (2018) 223201},
\href{http://arxiv.org/abs/1803.11166}{{\ttfamily arXiv:1803.11166 [hep-lat]}}.

\bibitem{Klco:2018kyo}
N.~Klco, E.~F. Dumitrescu, A.~J. McCaskey, T.~D. Morris, R.~C. Pooser, M.~Sanz,
  E.~Solano, P.~Lougovski, and M.~J. Savage, ``{Quantum-classical computation
  of Schwinger model dynamics using quantum computers},''
  \href{http://dx.doi.org/10.1103/PhysRevA.98.032331}{{\em Phys. Rev.}
  {\bfseries A98} no.~3, (2018) 032331},
\href{http://arxiv.org/abs/1803.03326}{{\ttfamily arXiv:1803.03326
  [quant-ph]}}.

\bibitem{Klco:2018zqz}
N.~Klco and M.~J. Savage, ``{Digitization of scalar fields for quantum
  computing},'' \href{http://dx.doi.org/10.1103/PhysRevA.99.052335}{{\em Phys.
  Rev.} {\bfseries A99} no.~5, (2019) 052335},
\href{http://arxiv.org/abs/1808.10378}{{\ttfamily arXiv:1808.10378
  [quant-ph]}}.

\bibitem{Lu:2018pjk}
H.-H. Lu {\em et~al.}, ``{Simulations of Subatomic Many-Body Physics on a
  Quantum Frequency Processor},''
  \href{http://dx.doi.org/10.1103/PhysRevA.100.012320}{{\em Phys. Rev.}
  {\bfseries A100} no.~1, (2019) 012320},
\href{http://arxiv.org/abs/1810.03959}{{\ttfamily arXiv:1810.03959
  [quant-ph]}}.

\bibitem{Klco:2019xro}
N.~Klco and M.~J. Savage, ``{Minimally-Entangled State Preparation of Localized
  Wavefunctions on Quantum Computers},''
\href{http://arxiv.org/abs/1904.10440}{{\ttfamily arXiv:1904.10440
  [quant-ph]}}.

\bibitem{Lamm:2018siq}
H.~Lamm and S.~Lawrence, ``{Simulation of Nonequilibrium Dynamics on a Quantum
  Computer},'' \href{http://dx.doi.org/10.1103/PhysRevLett.121.170501}{{\em
  Phys. Rev. Lett.} {\bfseries 121} no.~17, (2018) 170501},
\href{http://arxiv.org/abs/1806.06649}{{\ttfamily arXiv:1806.06649
  [quant-ph]}}.

\bibitem{Gustafson:2019mpk}
E.~Gustafson, Y.~Meurice, and J.~Unmuth-Yockey, ``{Quantum simulation of
  scattering in the quantum Ising model},''
  \href{http://dx.doi.org/10.1103/PhysRevD.99.094503}{{\em Phys. Rev.}
  {\bfseries D99} no.~9, (2019) 094503},
\href{http://arxiv.org/abs/1901.05944}{{\ttfamily arXiv:1901.05944 [hep-lat]}}.

\bibitem{Klco:2019evd}
N.~Klco, J.~R. Stryker, and M.~J. Savage, ``{SU(2) non-Abelian gauge field
  theory in one dimension on digital quantum computers},''
\href{http://arxiv.org/abs/1908.06935}{{\ttfamily arXiv:1908.06935
  [quant-ph]}}.

\bibitem{Alexandru:2019ozf}
{\bfseries NuQS} Collaboration, A.~Alexandru, P.~F. Bedaque, H.~Lamm, and
  S.~Lawrence, ``{$\sigma$ Models on Quantum Computers},''
  \href{http://dx.doi.org/10.1103/PhysRevLett.123.090501}{{\em Phys. Rev.
  Lett.} {\bfseries 123} no.~9, (2019) 090501},
\href{http://arxiv.org/abs/1903.06577}{{\ttfamily arXiv:1903.06577 [hep-lat]}}.

\bibitem{Alexandru:2019nsa}
{\bfseries NuQS} Collaboration, A.~Alexandru, P.~F. Bedaque, S.~Harmalkar,
  H.~Lamm, S.~Lawrence, and N.~C. Warrington, ``{Gluon Field Digitization for
  Quantum Computers},''
  \href{http://dx.doi.org/10.1103/PhysRevD.100.114501}{{\em Phys. Rev.}
  {\bfseries D100} no.~11, (2019) 114501},
\href{http://arxiv.org/abs/1906.11213}{{\ttfamily arXiv:1906.11213 [hep-lat]}}.

\bibitem{Mueller:2019qqj}
N.~Mueller, A.~Tarasov, and R.~Venugopalan, ``{Deeply inelastic scattering
  structure functions on a hybrid quantum computer},''
\href{http://arxiv.org/abs/1908.07051}{{\ttfamily arXiv:1908.07051 [hep-th]}}.

\bibitem{Lamm:2019uyc}
{\bfseries NuQS} Collaboration, H.~Lamm, S.~Lawrence, and Y.~Yamauchi,
  ``{Parton Physics on a Quantum Computer},''
\href{http://arxiv.org/abs/1908.10439}{{\ttfamily arXiv:1908.10439 [hep-lat]}}.

\bibitem{Magnifico:2019kyj}
G.~Magnifico, M.~Dalmonte, P.~Facchi, S.~Pascazio, F.~V. Pepe, and
  E.~Ercolessi, ``{Real Time Dynamics and Confinement in the $Z_{n}$
  Schwinger-Weyl lattice model for 1+1 QED},''
\href{http://arxiv.org/abs/1909.04821}{{\ttfamily arXiv:1909.04821
  [quant-ph]}}.

\bibitem{Chakraborty:2019}
B.~Chakraborty, M.~Honda, T.~Izubuchi, Y.~Kikuchi, and A.~Tomiya, ``{Digital
  Quantum Simulation of the Schwinger Model with Topological Term via Adiabatic
  State Preparation},'' \href{http://arxiv.org/abs/2001.00485}{{\ttfamily
  arXiv:2001.00485 [hep-lat]}}.

\bibitem{Kharzeev:2020kgc}
D.~E. Kharzeev and Y.~Kikuchi, ``{Real-time chiral dynamics from a digital
  quantum simulation},''
  \href{http://dx.doi.org/10.1103/PhysRevResearch.2.023342}{{\em Phys. Rev.
  Res.} {\bfseries 2} no.~2, (2020) 023342},
  \href{http://arxiv.org/abs/2001.00698}{{\ttfamily arXiv:2001.00698
  [hep-ph]}}.

\bibitem{Shaw:2020udc}
A.~F. Shaw, P.~Lougovski, J.~R. Stryker, and N.~Wiebe, ``{Quantum Algorithms
  for Simulating the Lattice Schwinger Model},''
  \href{http://arxiv.org/abs/2002.11146}{{\ttfamily arXiv:2002.11146
  [quant-ph]}}.

\bibitem{sahinoslu2020hamiltonian}
B.~Şahinoğlu and R.~D. Somma, ``Hamiltonian simulation in the low energy
  subspace,'' \href{http://arxiv.org/abs/2006.02660}{{\ttfamily
  arXiv:2006.02660 [quant-ph]}}.

\bibitem{Paulson:2020zjd}
D.~Paulson {\em et~al.}, ``{Towards simulating 2D effects in lattice gauge
  theories on a quantum computer},''
  \href{http://arxiv.org/abs/2008.09252}{{\ttfamily arXiv:2008.09252
  [quant-ph]}}.

\bibitem{Mathis:2020fuo}
S.~V. Mathis, G.~Mazzola, and I.~Tavernelli, ``{Toward scalable simulations of
  Lattice Gauge Theories on quantum computers},''
  \href{http://dx.doi.org/10.1103/PhysRevD.102.094501}{{\em Phys. Rev. D}
  {\bfseries 102} no.~9, (2020) 094501},
  \href{http://arxiv.org/abs/2005.10271}{{\ttfamily arXiv:2005.10271
  [quant-ph]}}.

\bibitem{Ji:2020kjk}
{\bfseries NuQS} Collaboration, Y.~Ji, H.~Lamm, and S.~Zhu, ``{Gluon Field
  Digitization via Group Space Decimation for Quantum Computers},''
  \href{http://arxiv.org/abs/2005.14221}{{\ttfamily arXiv:2005.14221
  [hep-lat]}}.

\bibitem{Raychowdhury:2019iki}
I.~Raychowdhury and J.~R. Stryker, ``{Loop, string, and hadron dynamics in
  SU(2) Hamiltonian lattice gauge theories},''
  \href{http://dx.doi.org/10.1103/PhysRevD.101.114502}{{\em Phys. Rev. D}
  {\bfseries 101} no.~11, (2020) 114502},
  \href{http://arxiv.org/abs/1912.06133}{{\ttfamily arXiv:1912.06133
  [hep-lat]}}.

\bibitem{Davoudi:2020yln}
Z.~Davoudi, I.~Raychowdhury, and A.~Shaw, ``{Search for Efficient Formulations
  for Hamiltonian Simulation of non-Abelian Lattice Gauge Theories},''
  \href{http://arxiv.org/abs/2009.11802}{{\ttfamily arXiv:2009.11802
  [hep-lat]}}.

\bibitem{Dasgupta:2020itb}
R.~Dasgupta and I.~Raychowdhury, ``{Cold Atom Quantum Simulator for String and
  Hadron Dynamics in Non-Abelian Lattice Gauge Theory},''
  \href{http://arxiv.org/abs/2009.13969}{{\ttfamily arXiv:2009.13969
  [hep-lat]}}.

\bibitem{Magnifico:2018wek}
G.~Magnifico, D.~Vodola, E.~Ercolessi, S.~P. Kumar, M.~M\"uller, and
  A.~Bermudez, ``{Symmetry-protected topological phases in lattice gauge
  theories: topological QED$_2$},''
  \href{http://dx.doi.org/10.1103/PhysRevD.99.014503}{{\em Phys. Rev. D}
  {\bfseries 99} no.~1, (2019) 014503},
  \href{http://arxiv.org/abs/1804.10568}{{\ttfamily arXiv:1804.10568
  [cond-mat.quant-gas]}}.

\bibitem{PhysRevD.107.L071502}
K.~Ikeda, ``Criticality of quantum energy teleportation at phase transition
  points in quantum field theory,''
  \href{http://dx.doi.org/10.1103/PhysRevD.107.L071502}{{\em Phys. Rev. D}
  {\bfseries 107} (Apr, 2023) L071502}.
  \url{https://link.aps.org/doi/10.1103/PhysRevD.107.L071502}.

\bibitem{Feynman1982}
R.~P. Feynman, ``Simulating physics with computers,''
  \href{http://dx.doi.org/10.1007/BF02650179}{{\em International Journal of
  Theoretical Physics} {\bfseries 21} no.~6, (Jun, 1982) 467--488}.
  \url{https://doi.org/10.1007/BF02650179}.

\bibitem{Preskill:2018fag}
J.~Preskill, ``{Simulating quantum field theory with a quantum computer},''
  \href{http://dx.doi.org/10.22323/1.334.0024}{{\em PoS} {\bfseries
  LATTICE2018} (2018) 024}, \href{http://arxiv.org/abs/1811.10085}{{\ttfamily
  arXiv:1811.10085 [hep-lat]}}.

\bibitem{Jordan:2012xnu}
S.~P. Jordan, K.~S.~M. Lee, and J.~Preskill, ``{Quantum Algorithms for Quantum
  Field Theories},'' \href{http://dx.doi.org/10.1126/science.1217069}{{\em
  Science} {\bfseries 336} (2012) 1130--1133},
  \href{http://arxiv.org/abs/1111.3633}{{\ttfamily arXiv:1111.3633
  [quant-ph]}}.

\bibitem{Preskill:2021apy}
J.~Preskill, ``{Quantum computing 40 years later},''
  \href{http://arxiv.org/abs/2106.10522}{{\ttfamily arXiv:2106.10522
  [quant-ph]}}.

\bibitem{Jiang2012}
H.-C. Jiang, Z.~Wang, and L.~Balents, ``Identifying topological order by
  entanglement entropy,'' \href{http://dx.doi.org/10.1038/nphys2465}{{\em
  Nature Physics} {\bfseries 8} no.~12, (Dec, 2012) 902--905}.
  \url{https://doi.org/10.1038/nphys2465}.

\bibitem{PhysRevB.81.064439}
F.~Pollmann, A.~M. Turner, E.~Berg, and M.~Oshikawa, ``Entanglement spectrum of
  a topological phase in one dimension,''
  \href{http://dx.doi.org/10.1103/PhysRevB.81.064439}{{\em Phys. Rev. B}
  {\bfseries 81} (Feb, 2010) 064439}.
  \url{https://link.aps.org/doi/10.1103/PhysRevB.81.064439}.

\bibitem{Ikeda:2020agk}
K.~Ikeda, D.~E. Kharzeev, and Y.~Kikuchi, ``{Real-time dynamics of Chern-Simons
  fluctuations near a critical point},''
  \href{http://dx.doi.org/10.1103/PhysRevD.103.L071502}{{\em Phys. Rev. D}
  {\bfseries 103} no.~7, (2021) L071502},
  \href{http://arxiv.org/abs/2012.02926}{{\ttfamily arXiv:2012.02926
  [hep-ph]}}.

\bibitem{Kogut:1974ag}
J.~B. Kogut and L.~Susskind, ``{Hamiltonian Formulation of Wilson's Lattice
  Gauge Theories},'' \href{http://dx.doi.org/10.1103/PhysRevD.11.395}{{\em
  Phys. Rev.} {\bfseries D11} (1975) 395--408}.

\bibitem{Susskind:1976jm}
L.~Susskind, ``{Lattice Fermions},''
  \href{http://dx.doi.org/10.1103/PhysRevD.16.3031}{{\em Phys. Rev.} {\bfseries
  D16} (1977) 3031--3039}.

\bibitem{Florio:2023dke}
A.~Florio, D.~Frenklakh, K.~Ikeda, D.~Kharzeev, V.~Korepin, S.~Shi, and K.~Yu,
  ``{Real-time non-perturbative dynamics of jet production: quantum
  entanglement and vacuum modification},''
  \href{http://arxiv.org/abs/2301.11991}{{\ttfamily arXiv:2301.11991
  [hep-ph]}}.

\bibitem{Jordan:1928wi}
P.~Jordan and E.~P. Wigner, ``{About the Pauli exclusion principle},''
\href{http://dx.doi.org/10.1007/BF01331938}{{\em Z. Phys.} {\bfseries 47}
  (1928) 631--651}.

\bibitem{Filippone:2011aa}
M.~Filippone, S.~Dusuel, and J.~Vidal, ``Quantum phase transitions in fully
  connected spin models: an entanglement perspective,''
  \href{http://dx.doi.org/10.1103/PhysRevA.83.022327}{{\em Phys. Rev. A}
  {\bfseries 83} (2011) 022327},
  \href{http://arxiv.org/abs/1101.3654}{{\ttfamily 1101.3654}}.
  \url{https://arxiv.org/pdf/1101.3654.pdf}.

\bibitem{PhysRevA.95.042321}
Y.~Susa, J.~F. Jadebeck, and H.~Nishimori, ``Relation between quantum
  fluctuations and the performance enhancement of quantum annealing in a
  nonstoquastic hamiltonian,''
  \href{http://dx.doi.org/10.1103/PhysRevA.95.042321}{{\em Phys. Rev. A}
  {\bfseries 95} (Apr, 2017) 042321}.
  \url{https://link.aps.org/doi/10.1103/PhysRevA.95.042321}.

\bibitem{Fischler:1978ms}
W.~Fischler, J.~B. Kogut, and L.~Susskind, ``{Quark Confinement in Unusual
  Environments},'' \href{http://dx.doi.org/10.1103/PhysRevD.19.1188}{{\em Phys.
  Rev. D} {\bfseries 19} (1979) 1188}.

\bibitem{Kao:1994wv}
Y.-C. Kao and Y.-W. Lee, ``{Inhomogeneous chiral condensate in the Schwinger
  model at finite density},''
  \href{http://dx.doi.org/10.1103/PhysRevD.50.1165}{{\em Phys. Rev. D}
  {\bfseries 50} (1994) 1165--1166}.

\bibitem{Nagy:2004ey}
S.~Nagy, J.~Polonyi, and K.~Sailer, ``{Periodic ground state for the charged
  massive Schwinger model},''
  \href{http://dx.doi.org/10.1103/PhysRevD.70.105023}{{\em Phys. Rev. D}
  {\bfseries 70} (2004) 105023},
  \href{http://arxiv.org/abs/hep-th/0405156}{{\ttfamily arXiv:hep-th/0405156}}.

\bibitem{Calabrese:2009qy}
P.~Calabrese and J.~Cardy, ``{Entanglement entropy and conformal field
  theory},'' \href{http://dx.doi.org/10.1088/1751-8113/42/50/504005}{{\em J.
  Phys. A} {\bfseries 42} (2009) 504005},
  \href{http://arxiv.org/abs/0905.4013}{{\ttfamily arXiv:0905.4013
  [cond-mat.stat-mech]}}.

\bibitem{Sakai:2004cn}
T.~Sakai and S.~Sugimoto, ``{Low energy hadron physics in holographic QCD},''
  \href{http://dx.doi.org/10.1143/PTP.113.843}{{\em Prog. Theor. Phys.}
  {\bfseries 113} (2005) 843--882},
  \href{http://arxiv.org/abs/hep-th/0412141}{{\ttfamily arXiv:hep-th/0412141}}.

\bibitem{yee2011holographic}
H.-U. Yee and I.~Zahed, ``Holographic two dimensional qcd and chern-simons
  term,'' {\em Journal of High Energy Physics} {\bfseries 2011} no.~7, (2011)
  1--19.

\bibitem{goldstein2018symmetry}
M.~Goldstein and E.~Sela, ``Symmetry-resolved entanglement in many-body
  systems,'' {\em Physical review letters} {\bfseries 120} no.~20, (2018)
  200602.

\end{thebibliography}\endgroup

\begin{appendix}
\clearpage
\begin{widetext}
\section{Central Charge Determination}
\begin{figure*}[!htbp]
    \centering
    \includegraphics[width=0.4\textwidth]{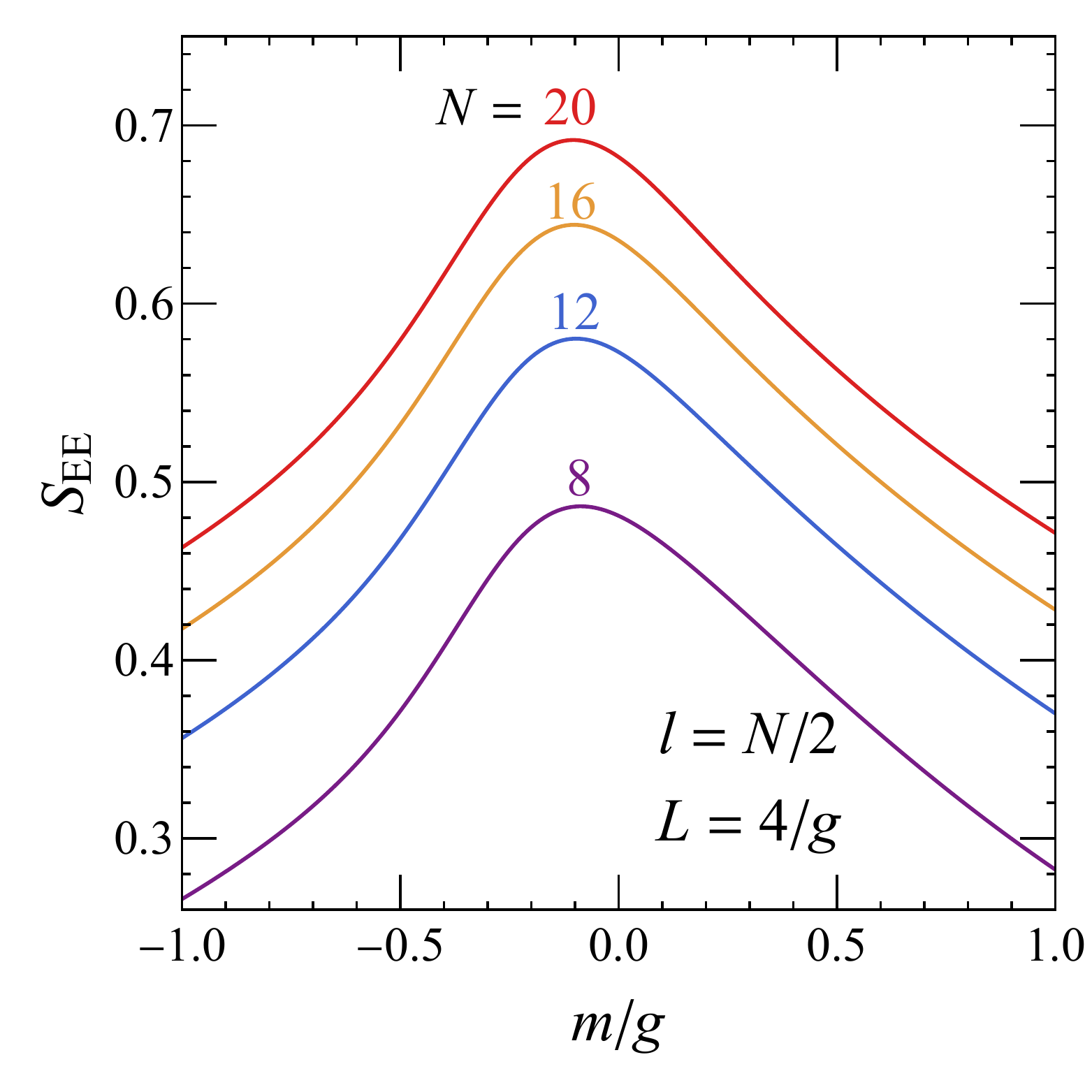}
    \includegraphics[width=0.4\textwidth]{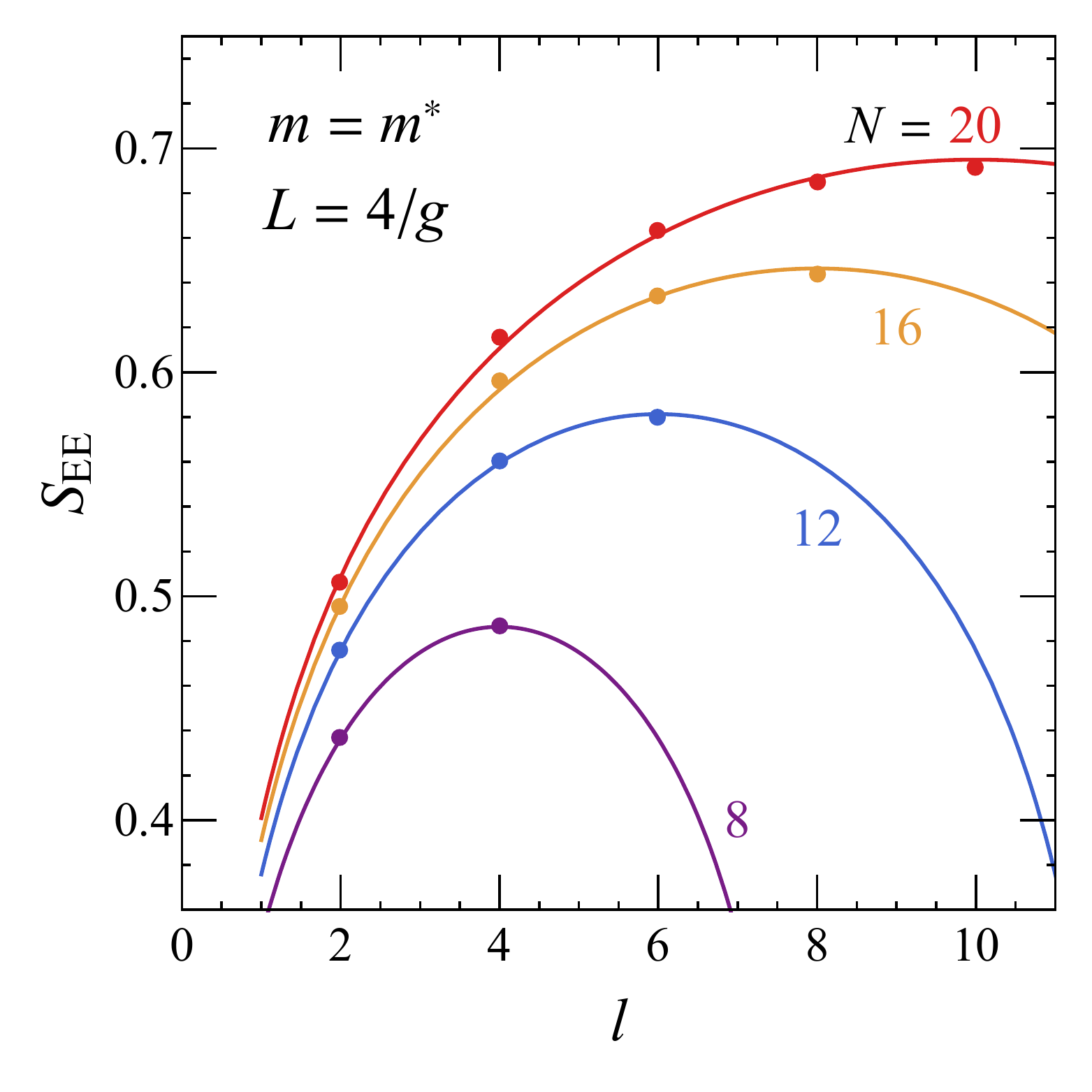}
    \caption{Mass(left) and length(right) dependence of entanglement entropy. Curves in the middle panel are fitted results using Eq.~\protect{\eqref{eq:SEE_l}}. Purple, blue, orange, and red curves correspond to total number of lattice site $N=8$, $12$, $16$, and $20$, respectively. 
    Results are evaluated at vanishing chemical potential.}
    \label{fig:EE}
\end{figure*}
Here we supplement more details for Fig.~5 of the main text, where we present the result of central charge ($c$) at various lattice spacing ($a$) and number of sites ($N$). We take a system with length $L=4/g$ as an example.
For each choice of $a$ and $N$, we first determine the critical mass $(m^*)$ by maximizing $S_{EE}$ when $l=N/2$; see Fig.~\ref{fig:EE} (left). Then, we take $m = m^*$ and compute $S_{EE,l}$ at different $l$'s, which are presented as dots in Fig.~\ref{fig:EE} (right). Finally, the $\ell$ dependence of $S_{EE}$ is fitted according to Eq.~\protect{\eqref{eq:SEE_l} of the main text, and the best fit results are also shown in Fig.~\ref{fig:EE}~(right)}.
\end{widetext}
\end{appendix}

\end{document}